\documentclass{aa}
\usepackage[usenames,dvipsnames]{xcolor}
\usepackage[T1]{fontenc}
\usepackage{natbib}
\usepackage{csquotes}
\usepackage{xspace}
\usepackage{amssymb}
\usepackage{amsmath}
\usepackage{graphicx}
\usepackage{threeparttable}
\usepackage{hyperref}
\usepackage{siunitx}
\usepackage{tikz}
\usepackage{subcaption}
\usepackage{aas_macros}

\usetikzlibrary{arrows.meta}
\tikzset{%
  >={Latex[width=2mm,length=2mm]},
            base/.style = {rectangle, rounded corners, draw=black,
                           minimum width=2cm, minimum height=0.5cm,
                           text centered, font=\sffamily},
  activityStarts/.style = {base, fill=blue!30},
       startstop/.style = {base, fill=red!30},
    activityRuns/.style = {base, fill=green!30},
         process/.style = {base, minimum width=2.5cm, fill=orange!15,
                           font=\ttfamily},
}

\DeclareSIUnit\erg{erg}
\DeclareSIUnit\Myr{Myr}
\DeclareSIUnit\AU{AU}

\def\erg{\hbox{erg}}

\def\cm{\hbox{cm}}

\def\AU{\hbox{AU}}

\def\Omegak{\ensuremath{\Omega_\mathrm{K}}\xspace}

\begin{document}

\title{How drifting and evaporating pebbles shape giant planets \\ II: Volatiles and refractories in atmospheres
}  \titlerunning{volatiles and refractories in giant planet atmospheres}
\authorrunning{Schneider \& Bitsch}
\author{Aaron David~Schneider$^{1,2,3}$ \& Bertram~Bitsch$^{1}$}
\institute{
  (1) Max-Planck-Institut f\"ur Astronomie, K\"onigstuhl 17, 69117 Heidelberg, Germany\\
  (2) Centre for ExoLife Sciences, Niels Bohr Institute, Øster Voldgade 5, 1350 Copenhagen, Denmark\\
  (3) Instituut voor Sterrenkunde, KU Leuven, Celestijnenlaan 200D, B-3001 Leuven, Belgium
} \date{\today}
\offprints{B. Bitsch,\\ \email{bitsch@mpia.de}}

\abstract{Upcoming studies of extrasolar gas giants will give precise insights into the composition of planetary atmospheres, with the ultimate goal of linking it to the formation history of the planet. Here, we investigate how drifting and evaporating pebbles that enrich the gas phase of the disk influence the chemical composition of growing and migrating gas giants. To achieve this goal, we perform semi-analytical 1D models of protoplanetary disks, including viscous evolution, pebble drift, and evaporation, to simulate the growth of planets from planetary embryos to Jupiter-mass objects by the accretion of pebbles and gas while they migrate through the disk. The gas phase of the protoplanetary disk is enriched due to the evaporation of inward drifting pebbles crossing evaporation lines, leading to the accretion of large amounts of volatiles into the planetary atmosphere. As a consequence, gas-accreting planets are enriched in volatiles (C, O, N) compared to refractories (e.g., Mg, Si, Fe) by up to a factor of 100, depending on the chemical species, its exact abundance and volatility, and the disk's viscosity. A simplified model for the formation of Jupiter reveals that its nitrogen content can be explained by inward diffusing nitrogen-rich vapor, implying that Jupiter did not need to form close to the N$_2$ evaporation front as indicated by previous simulations. However, our model predicts an excessively low oxygen abundance for Jupiter, implying either Jupiter's migration across the water ice line (as in the grand tack scenario) or an additional accretion of solids into the atmosphere (which can also increase Jupiter's carbon abundance, ultimately changing the planetary C/O ratio). The accretion of solids, on the other hand, will increase the refractory-to-volatile ratio in planetary atmospheres substantially. We thus conclude that the volatile-to-refractory ratio in planetary atmospheres can place a strong constraint on planet formation theories (in addition to elemental ratios), especially on the amount of solids accreted into atmospheres, making it an important target for future observations.
}
\maketitle


\section{Introduction}

Even though more than 4000 exoplanets have been discovered so far \citep{NASA_exoplanet_archive}, it is still unclear how exactly these planets formed. In the core accretion scenario, the planetary core is built by the accretion of planetesimals \citep{IdaLinI,Alibert2005,IdaLinIV,IdaLinV,IdaLinVI,Mordasini2012,Alibert2013,Emsenhuber2020} or pebbles \citep{Lambrechts2012, Lambrechts2014, Bitsch2015, Ali-Dib2017, Ndugu2018, Bruegger2018, Lambrechts2019, Bitsch2019}. While the underlying mechanisms of pebble and planetesimal accretion are fundamentally different, both models require a sufficiently fast formation of a planetary core of a few Earth masses before gas accretion can begin \citep{Pollack1996, Ikoma2000}. 

Hope to constrain the planet formation pathway is based on the detailed characterizations of the planetary atmosphere, especially the C/O ratio \citep[e.g.,][]{Oeberg2011}. Previous observations of planetary C/O ratios have revealed super-solar C/O values \citep{Brewer2017}, indicating that these planets might originate from beyond the water ice line, where the C/O ratio in the disk is larger \citep{Oeberg2011, Madhusudhan2017, Booth2017_pebble_evap, Notsu2020, Schneider2021I}. Furthermore, current spectroscopic observations of WASP-121b have revealed the absence of VO and TiO \citep{Merrit2020} and the presence of neutral transition metals such as vanadium \citep{Ben-Yami2020}, providing further constraints on planet formation. The upcoming {\it James Webb Space Telescope} (JWST) and the {\it Atmospheric Remote-sensing Infrared Exoplanet Large-survey} (ARIEL) missions are expect to constrain planetary C/O ratios and other atmospheric elemental abundances for many more targets, building a large sample that can be used to constrain planet formation theories.

In \citet{Schneider2021I} we investigate the origin of the total heavy element content of giant planets, where the evaporation of drifting pebbles that pollute the disk gas with heavy elements (see also \citealt{Banzatti2020, Zhang2020}) can account for the large fraction of heavy elements inside the giant planets \citep{Thorngren2016}. As a consequence, this process would significantly enrich the planetary atmosphere with volatile elements (e.g., C, O, N) but leave the planetary atmosphere devoid of refractory elements (e.g., Mg, Si, Fe).

On the other hand, \citet{Owen1999} and \citet{Atreya2016} proposed that the super-solar abundances of carbon, nitrogen, and sulfur in the atmosphere of Jupiter could be explained by the accretion of planetesimals. Furthermore, \citet{Oeberg2019} and \citet{Bosman2019} suggest that Jupiter's nitrogen abundance is most likely explained by the formation of Jupiter beyond the N$_2$ evaporation line, allowing the accretion of nitrogen-rich solids. However, planetesimals and comets are, to a very large fraction, made out of refractory elements. Accreting planetesimals into planetary atmospheres would thus result in a large fraction of refractory elements inside these atmospheres as well.

It is clear that the enrichment of gas giant atmospheres via vapor-enriched gas and planetesimals is fundamentally different and would also lead to different atmospheric compositions. In this work we use our previous model of pebble drift and evaporation \citep{Schneider2021I} to study the detailed chemical composition of gas giants, with a special emphasis on volatile and refractory elements as well as on the planetary C/O ratio. We first focus on exoplanet formation and then show results of a simplified model for the inferred composition of Jupiter and Saturn.
\section{Methods}\label{sec:methods}
We modeled the formation of gas giant planets by the accretion of pebbles and gas in viscously evolving disks while tracing the chemical composition of the migrating planet. The newly developed code \texttt{chemcomp} used in these simulations is explained in detail in \citet{Schneider2021I}. These simulations include the evaporation of inward drifting pebbles and the condensation of outward diffusing gas at evaporation lines. Within these models we simulated the growth and migration of single planetary embryos all the way to gas giants.  

The initial solid surface density was 
\begin{equation}
        \Sigma_\mathrm{Z} = \epsilon_0 \Sigma_\mathrm{gas},
\end{equation}
where $\epsilon_0$ is the solid-to-gas ratio and $\Sigma_\mathrm{gas}$ is the gas surface density. The solid-to-gas ratio depends on the position in the disk, where a higher solid-to-gas ratio is used when volatile species are frozen out \citep[see][for more details]{Schneider2021I}. For all our simulations, we set $\epsilon_0 = 1.5\%$ in the outer disk. The gas surface density can be calculated from the disk mass ($M_0$) and disk radius ($R_0$). We utilized here the alpha-viscosity \citep{Shakura1973} prescription, which relates the viscosity to the numerical parameter ($\alpha$) that describes the turbulent strength. The viscosity is then given by 
\begin{equation}
        \nu = \alpha \frac{c_\mathrm{s}^2}{\Omegak},
\end{equation}
where $c_s$ is the isothermal sound speed and $\Omega_K$ the orbital period. 

Our pebble growth and evolution model is based on that in \citet{Birnstiel2012}, where the maximal grain size in the fragmentation-limited regime is determined by the disk viscosity as well as by the dust fragmentation velocity, which we set to $u_{\rm frag}=5$m/s, following laboratory constraints \citep{Gundlach2015}. We used a variable pebble density, $\rho_\bullet$, where the pebble density depends on the volatile-to-refractory ratio in the solid surface density \citep{Drazkowska2017}. The disk model was calculated on a logarithmically spaced grid of $N_{\rm Grid}=500$ grid cells in between $r_{\rm in}=0.1$ AU and $r_{\rm out}=1000$ AU. 

We assumed that the initial chemical composition of the protoplanetary disk is similar to the composition of the host star, for which we used the solar abundances (denoted as [Fe/H]=0) from \citet{Asplund2009}. We followed here the chemical compositions as outlined in \citet{Schneider2021I}. 

The Solar System provides evidence for the initial abundances not only from the solar photosphere, but also from meteorites. In particular, CI chondrites should have accreted refractory carbon grains. By calculating the ratio of carbon inside CI chondrites to carbon in the solar photosphere, we can roughly determine the amount of refractory carbon for our chemical models. Using the data from \citet{Lodders2003}, we find that roughly 10\% of the carbon should be in refractory components and the remaining carbon in volatile form. Using data from comets \citep{Altwegg2020Comet} results in a slightly higher refractory carbon content of around 20\%. Our standard chemical model therefore contains 20\% in refractory carbon, but we also investigated a situation where 60\% of all carbon is in refractory form (Table~\ref{tab:comp}), representing the refractory carbon fraction in the interstellar medium (ISM; e.g., \citealt{Bergin2015})\footnote{A lot of studies try to link the ISM refractory carbon abundances to the carbon abundance on Earth (e.g., \citealt{Klarmann2018, vantHoff2020, Li2021}), showing that there is clearly some uncertainty regarding the incorporation of carbon into solids.}. As more carbon grains are bound in refractories, less oxygen is bound in CO and CO$_2$, leaving more oxygen available to form water, in turn increasing the water abundance within our model (see Appendix~\ref{sec:disk}).

Applying the same argument for sulfur and nitrogen reveals that nearly all sulfur should be in refractory form, while nearly all nitrogen should be in volatile form \citep{Lodders2003}, in line with our chemical model \citep{Schneider2021I}. Of course, this argument is based on the assumption that CI chondrites accreted the full refractory component of the corresponding species. 

\begin{table*}
\centering
\caption[Condensation temperatures]{Condensation temperatures and volume mixing ratios.}
\begin{tabular}{c c c c}
\hline\hline
Species (Y) & $T_{\text{cond}}$ {[}K{]} & $v_{\text{Y, 20\% C}}$ & $v_{\text{Y, 60\% C}}$\\ \hline 
CO & 20  & 0.45 $\times$ C/H &  0.2 $\times$ C/H  \\
CH$_4$ & 30 & 0.25 $\times$ C/H  & 0.1 $\times$ C/H \\
CO$_2$ & 70 & 0.1 $\times$ C/H & 0.1 $\times$ C/H \\
C (carbon grains) & 631 & 0.2 $\times$ C/H & 0.6 $\times$ C/H\\ \hline
\end{tabular}
\begin{tablenotes}
        \item \textbf{Notes:} The table displays only the carbon-bearing species used in our work. The full condensation sequence of our chemical model is shown in \citet{Schneider2021I}. 
\end{tablenotes}
\label{tab:comp}
\end{table*}

\begin{table}
        \caption{Parameters used throughout this paper.}
        \begin{subtable}{.2\textwidth}
        \vspace{0pt}
        \centering
        \begin{tabular}{c c}
                \hline\hline
                Quantity & Value\\
                \hline
                $a_{p,0}$                        & $(3,10,30)$ AU\\

                $t_0$                        & \SI{0.05}{Myr}\\
                $\kappa_\mathrm{env}$& $\SI{0.05}{\cm\squared\per\g}$\\
                \hline
        \end{tabular}
        \caption{Planet}
        \begin{tabular}{c c}
                \hline\hline
                Quantity & Value\\
                \hline
                $r_\mathrm{in}$       & $\SI{0.1}{AU}$\\
                $r_\mathrm{out}$      & $\SI{1000}{AU}$\\
                $N_\mathrm{Grid}$     & $\num{500}$\\
                \hline
        \end{tabular}
        \caption{Grid}
        \end{subtable}\hfill
        \begin{subtable}{.25\textwidth}
        \vspace{0pt}
                \centering
                \begin{tabular}{c c}
                \hline\hline
                Quantity & Value\\
                \hline
                $\alpha$           & $(1,5,10)\times10^{-4}$\\
                $\alpha_z$         & $\num{1e-4}$\\
                $M_0$              & $\SI{0.128}{M_\odot}$\\
                $R_0$              & $\SI{137}{AU}$\\
                $[\mathrm{Fe}/\mathrm{H}]$ & 0\\
                $t_\mathrm{evap}$  & $\SI{3}{Myr}$\\
                $\epsilon_0$       & $1.5\%$\\
                $u_\mathrm{frag}$  & $\SI{5}{\m\per\s}$\\
                \hline
        \end{tabular}
        \caption{Disk}
        \end{subtable}
        \begin{tablenotes}
                \item \textbf{Notes:} Parameters used for the initialization of \texttt{chemcomp} that are used throughout this paper, divided into planetary, numerical, and disk parameters. The detailed explanation of these parameters can be found in \citet{Schneider2021I}.
        \end{tablenotes}

        \label{tab:parameters}
\end{table}

The growth of the planets is divided into two phases. First, the planet accretes pebbles \citep[e.g.,][]{Johansen2017peb} until it reaches the pebble isolation mass \citep[e.g.,][]{Lambrechts2014, Bitsch2018_peb_iso}, where pebble accretion stops. During the solid accretion phase, we attributed 90\% of the solids to the core and 10\% of the solids to a primordial planetary atmosphere during core buildup, following the idea that pebbles evaporate during accretion \citep{Hori2011, Brouwers2020}. The pebble isolation mass is smaller in the inner regions of the disk, due to the flaring disk structure \citep[e.g.,][]{Chiang1997, Bitsch2015b}, resulting in smaller core masses of inner forming planets. The planet then starts to accrete gas from the protoplanetary disk, via a slow gas contraction followed by rapid runaway gas accretion, which is limited by the disk's supply rate, determined by viscosity. For gas accretion, we followed the approach outlined in \citet{Ndugu2021}, where the envelope contraction rate is modeled via \citet{Ikoma2000}, which depends not only on the planetary core mass, but also on the envelope opacity, $\kappa_{\rm env}$. For $\kappa_{\rm env}$ we used a constant value of 0.05 cm$^2$/g \citep{Movshovitz2008}.

We used the same set of parameters (see Table \ref{tab:parameters}) as that used in \citet{Schneider2021I} and only varied the initial position, $a_\mathrm{p,0}$, of the planet and the viscous $\alpha$ parameter. Throughout this work we compare the formation of planets in different disks that harbor different viscosities for the disk evolution and pebble growth (see Table \ref{tab:parameters}). However, the vertical distribution of the pebbles is assumed to happen with low turbulence ($\alpha_{\rm z} = 10^{-4}$ for all simulations).

Calculating the atmospheric composition within our model is mainly influenced by the initial assumption that once the planet reaches the pebble isolation mass, it exerts a pressure bump, blocking all available solids exterior to planet. As a consequence, the planets in our model can, at this stage, only accrete evaporated material (also known as vapor), not solids. This is further discussed in Sect.~\ref{sec:owen}.

\section{Planet formation}\label{sec:formation}

During the planetary growth, the planet migrates first in type-I migration and then in type-II migration once it opens a gap in the protoplanetary disk. The migration speed during type-II is directly proportional to the disk's viscosity, allowing planets forming in disks with high viscosities to migrate farther inward compared to planets forming in disks with low viscosities \citep{Baruteau2014}. We implanted the planet at t=0.05 Myr and stopped the integration either at the end of the disk's lifetime (at $t_\mathrm{evap}=\SI{3}{\Myr}$) or when the planet reaches 0.2 AU. 

Planetary embryos that start to grow at 3 AU initially first migrate outward due to the effects of the heating torque \citep{Benitez-Llambay2015, Masset2017, Baumann2020} but then migrate rapidly inward before they open deep gaps. This is caused by the relatively long envelope contraction phase for these small cores. In fact, the planet forming in a disk with $\alpha=10^{-3}$ migrates down to the disk's inner edge before 1 Myr of evolution because it is unable to open a deep gap, in contrast to the planets forming in lower viscosity environments. 

The planetary embryos starting at 10 and 30 AU evolve in a very similar way. Namely, the planets forming in disks with higher viscosities migrate inward more efficiently due to the delayed gap opening, and they grow to larger masses because the disk's gas supply rate is larger due to the larger viscosity. In fact, the planets forming in very low viscosity environments ($\alpha=10^{-4}$) only migrate inward for a few AU relative to their starting position. 

\begin{figure*}
        \centering
        \includegraphics[width=\textwidth]{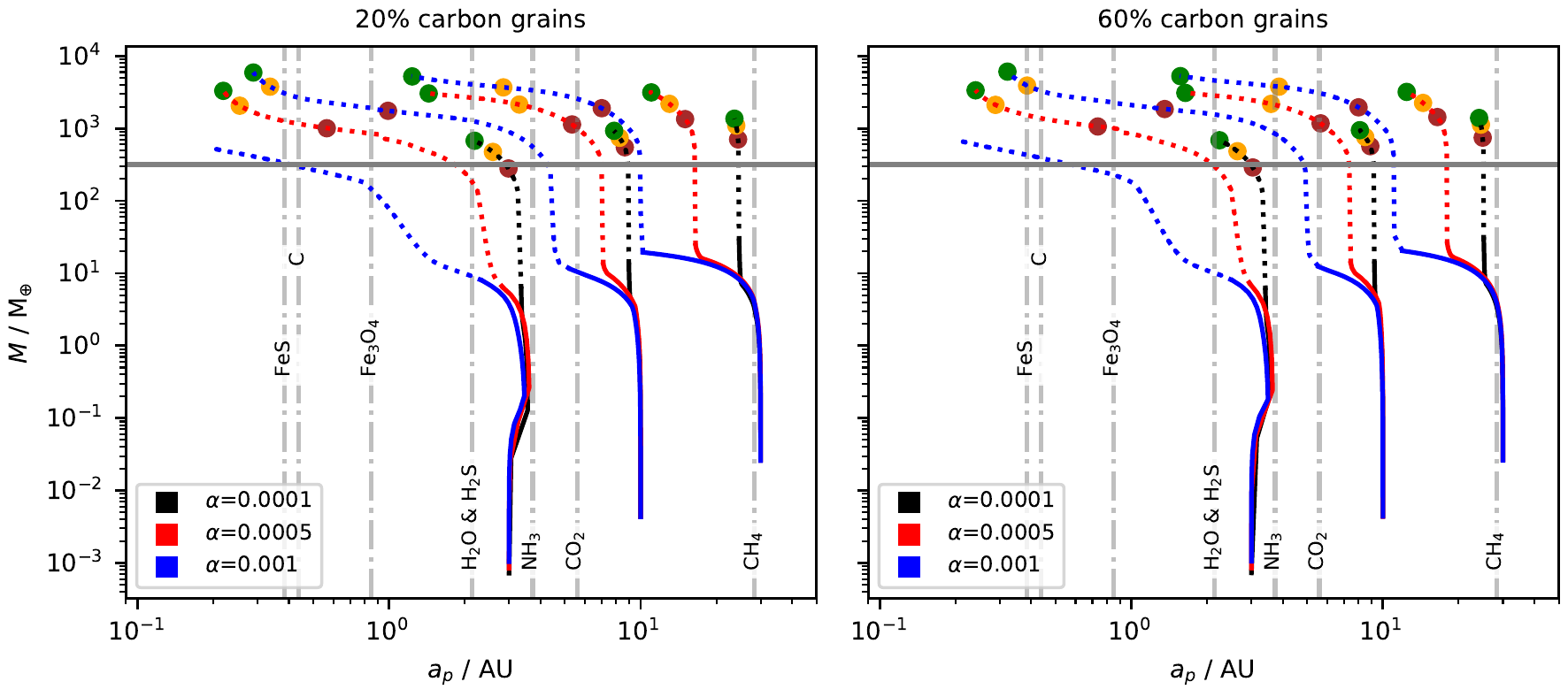}
        \caption{Growth tracks of planets that accrete pebbles (solid lines) and gas (dashed lines) starting at different initial positions in disks with different values for $\alpha$ (indicated by color). The dashed gray lines mark the positions of evaporation lines for $\alpha=\num{5e-4}$. The solid gray horizontal line marks the mass of Jupiter. The brown, yellow, and green dots mark an evolution time of $t=\SI{1}{Myr}$, $t=\SI{2}{Myr}$, and $t=\SI{3}{Myr}$, respectively. The planet starting at 3 AU in a disk with $\alpha=0.001$ reaches the inner edge of the disk before 1 Myr of evolution. The left panels display models with 20\% of the carbon abundance locked in refractory carbon grains, whereas the right panels show models with 60\% locked in carbon grains. Model parameters can be found in Table \ref{tab:parameters}.}
        \label{fig:growthtracks}
\end{figure*}

The inward migration speed of the gas giants reduces as the planets grow (upward shift in the growth tracks toward their end; Fig.~\ref{fig:growthtracks}). This is caused by the fact that the type-II migration rate also scales with the planetary mass, resulting in slower migration for more massive planets \citep{Baruteau2014}.

\section{Volatiles versus refractories}\label{sec:owen}

\begin{figure*}
        \centering
        \includegraphics[width=\textwidth]{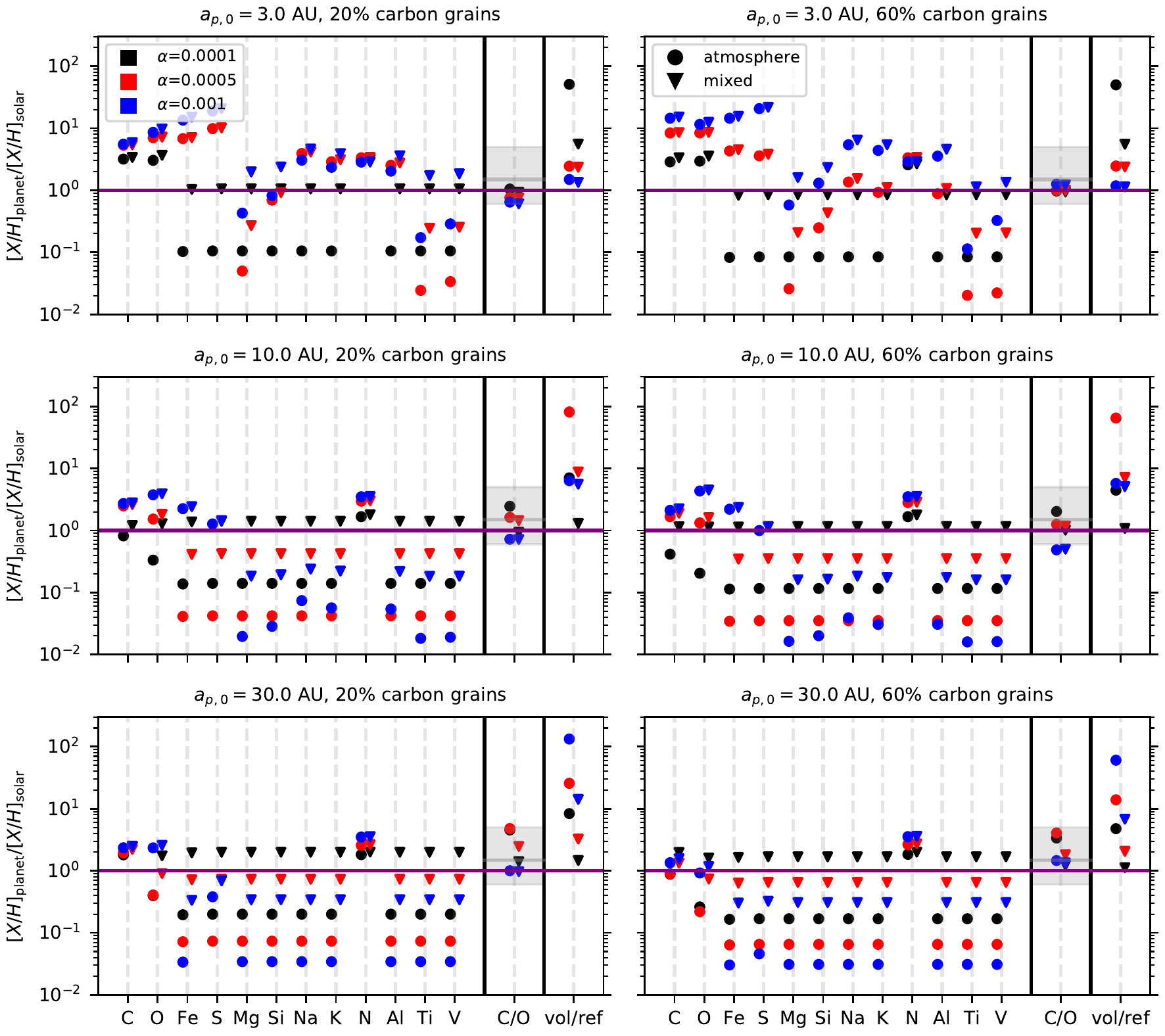}
        \caption{Final atmospheric compositions of the planets shown in Fig. \ref{fig:growthtracks} normalized to the solar composition. We also show the C/O ratio normalized to the solar value on the right of each panel, with Jupiter's C/O ratio marked in gray \citep{Atreya2016, Li2020Jup}, as well as the volatile-to-refractory ratio, which we calculate as the ratio of volatile to refractory molecules in the atmosphere. The left panels display models with 20\% of the carbon abundance locked in refractory carbon grains, whereas the right panels show models with 60\% locked in carbon grains. The circles mark the composition of the pure atmosphere, while the triangles mark the composition if the core were completely mixed into the atmosphere.}
        \label{fig:owen}        
\end{figure*}

In this section we discuss the elemental abundances in the planetary atmospheres of the planets formed in our model (see Fig.~\ref{fig:growthtracks}). Our planetary model does not include an atmospheric structure model, which would be needed for accurate simulations of the atmospheric composition gradients \citep[e.g.,][]{Vazan2018}. We thus assumed a perfectly mixed atmosphere, either with or without a complete mix of the planetary core into the atmosphere, showing the most extreme elemental abundances possible within our model. In the scope of this work, we define the volatile-to-refractory ratio as the ratio of volatile molecules (evaporation temperature below \SI{150}{K}) to refractory molecules (evaporation temperature above \SI{150}{K})

\subsection{Effects of pebble evaporation}

We show in Fig.~\ref{fig:owen} the elemental ratios inside the planetary atmosphere normalized to the solar composition for the pure atmosphere (circles) or when the core is completely mixed into the atmosphere (triangles). The advantage of this approach is that it shows the maximal (triangles) and minimal (circles) values of the atmospheric composition within our model.

Figure \ref{fig:owen} clearly shows an enrichment of volatile species (C, O, N) in the planetary atmospheres compared to refractory elements (e.g., Al, Mg, Si), leading to a high volatile-to-refractory ratio, as displayed. This is naturally explained by gas accretion, which allows the accretion of volatile-rich vapor (e.g., H$_2$O, CO$_2$, NH$_3$, CH$_4$, CO) but not of refractory elements, which are bound in pebbles blocked exterior to the planet. This potentially explains the absence of VO and TiO in WASP-121b \citep{Merrit2020}, although atmospheric effects such as transport and thermal evolution can also reduce the VO and TiO abundances \citep[e.g.,][]{Parmentier2016Clouds, Beatty2017}. This effect is also reflected by the decrease in the refractory content in the planetary atmosphere with increasing $\alpha$ values, where larger $\alpha$ values result in faster gas accretion and more massive planets. The accreted gas is refractory poor due to the blockage of refractory-rich pebbles exterior to the planet, which is caused by the pressure bump of the planet that can block pebbles from drifting inward. This dilutes the initially high refractory abundance in the atmosphere during core buildup\footnote{The planet forming at 3 AU in the disk with $\alpha=10^{-3}$ is an exception because it migrates to the inner edge of the disk before the end of the disk lifetime, where we stop accretion, resulting in the relatively high refractory abundance compared to the counterparts formed in low viscosity disks.}. 

One exception is the accretion of the FeS vapor of the planets migrating into the very inner, hot regions of the disk, where FeS evaporates and can be accreted in gaseous form. This results in super-solar Fe/H and S/H values in the atmospheres of the planets starting at 3 AU in disks with $\alpha > \num{1e-4}$. Our model assumes that a large fraction of sulfur is bound in FeS and only a small amount in H$_2$S \citep{Kama2019}. In contrast, comets in the Solar System seem to have a large H$_2$S abundance \citep{Flynn2006, Mumma2011}, hinting at a different sulfur distribution compared to our nominal model. On the other hand, recondensation of H$_2$S vapor at the H$_2$S evaporation front can locally increase the S/H ratio in the solids, potentially explaining the large H$_2$S abundance in comets. A similar effect at carbon-bearing evaporation fronts is invoked in \citet{Mousis2021} to explain the composition of the comet C/2016 R2. The other refractory species show different abundances for the planets migrating all the way to the inner disks because the different evaporation fronts are crossed at different times. In contrast, the refractory contents of the planets formed in the outer disk are similar for all elements because there the refractories are only in solid form and cannot be accreted with the gas in our model.

The inward drifting and evaporating pebbles enrich the gas in volatiles to super-solar values (C, O, N), explaining the super-solar values of volatile species in the planetary atmospheres. The exception here is oxygen, which can be subsolar for gas-accreting planets forming in the outer regions of the disk if they do not migrate across the CO$_2$ evaporation front, which would allow an efficient accretion of oxygen with the gas. This effect is clearly visible for the planets forming at 30 AU when comparing the oxygen abundance of the planet forming in a disk with $\alpha=10^{-3}$ and the planets forming in disks with lower viscosities (Fig.~\ref{fig:owen}).

Our simulations also show a clear trend regarding the planetary C/O ratio with increasing initial planetary position. Planets forming farther out have a larger C/O ratio compared to planets forming closer to the host star. This is a direct consequence of the C/O ratio in the gas phase of the protoplanetary disk, which increases with orbital distance due to the evaporation of carbon-bearing species \citep{Schneider2021I}.

If the planetary core is completely mixed into the planetary atmosphere, the atmosphere is enriched with the material from the core. In our simulations, this implies an enrichment with refractory species, which is clearly visible for all our simulations. Naturally, larger cores lead to more enrichment, where the core mass increases toward larger radii due to the flaring nature of the disk, increasing the pebble isolation mass. Volatile species, on the other hand, are only marginally increased, with only oxygen showing a major increase for the planets forming in the outer disk regions (due to the oxygen bound in CO$_2$ and water ice). As a consequence, the atmospheric C/O ratio is relatively unchanged, except for the planets forming in the outer disk, which show a drop in the C/O ratio due to the large oxygen content of the core. 

The model with 60\% carbon grains shows that the gas phase of the disk interior to the carbon grain evaporation line is enriched in carbon compared to the model with only 20\% carbon grains, while it is depleted beyond the carbon grain evaporation line. This means that planets that accrete most of the gas beyond the carbon grain evaporation line host atmospheres that have less carbon in a model with more carbon grains (see the right panels of Fig. \ref{fig:owen}) than those in the model with fewer carbon grains (see the left panels of Fig. \ref{fig:owen}). Contrary, planets that migrate all the way to the inner disk have enhanced carbon abundances in a model with more carbon grains. This clearly shows that a detailed chemical model is of crucial importance if the planet formation pathway is supposed to be constrained by C/O alone \citep[e.g.,][]{Notsu2020}. Nevertheless, the overall planetary C/O ratio is quite similar for planets formed within the two different chemical models.

Including more carbon grains in the chemical model (see Table \ref{tab:comp}) redistributes the carbon-rich evaporation fronts inward because of the reduced amount of CH$_4$ (see Appendix~\ref{sec:disk}). The enhanced evaporation front at \SI{631}{K} ($r\approx \SI{0.2}{AU}$) near the FeS evaporation line raises the C/O ratio of the gas phase C/O in the inner disk (see Fig. \ref{fig:sigma}), while lowering the pebble pileup and the pollution due to methane evaporation at the methane evaporation line. The other chemical elements are not affected by the change in the carbon distribution, and the planet formation (growth and migration times) remains largely unaffected as well.

Similar changes in the planetary composition are expected for other elements that can exist in volatile or refractory form (e.g., sulfur).

\subsection{Effects of additional solids}
\label{sec:solids}
\begin{table}
        \centering
        \caption{Additional solid enrichment.} 
        \begin{tabular}{c c c}
                \hline\hline
                & 20\% carbon grains & 60\% carbon grains\\
                \hline
                Fig.~\ref{fig:owen_pla_eff} & \SI{30}{M_\odot} & \SI{30}{M_\odot}\\
                Jupiter & \SI{9}{M_\odot} & \SI{12}{M_\odot}\\
                Saturn & \SI{11}{M_\odot} & \SI{14}{M_\odot}\\
                \hline
        \end{tabular}   
        \begin{tablenotes}
                \item \textbf{Notes:} Additional solid enrichment for Figs. \ref{fig:owen_pla_eff} and \ref{fig:JupSat}.
        \end{tablenotes}
        \label{tab:add_solids}
\end{table}

\begin{figure*}
        \centering
        \includegraphics[width=\textwidth]{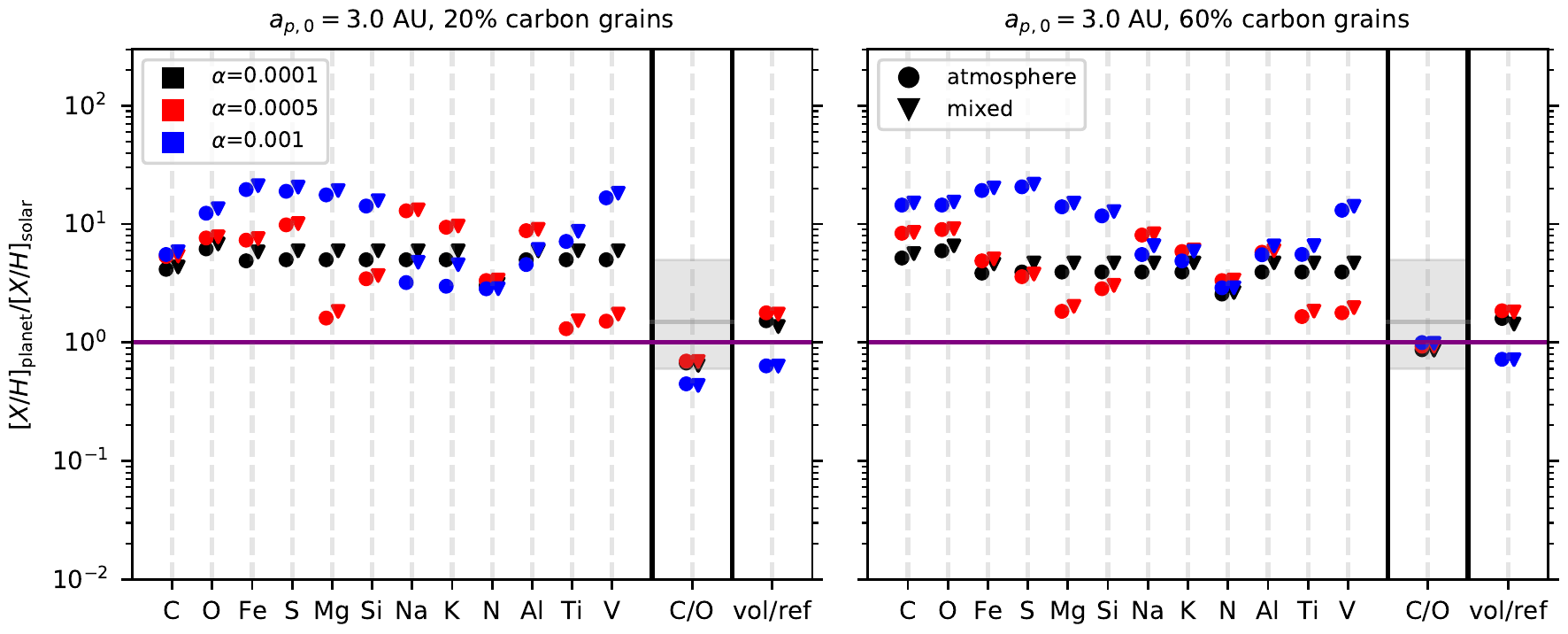}
        \caption{As Fig.~\ref{fig:owen}, but with the addition of 30 Earth masses of solids into the atmosphere. We only show the results of the planets starting at 3.0 AU because the addition of extra solids into the atmospheres results in the same trends for planets starting farther away from the star.}
        \label{fig:owen_pla_eff}        
\end{figure*}

Planetary atmospheres can also be enriched via collisions \citep{Ogihara2021} or via the accretion of planetesimals. This can happen either during the buildup of the planetary atmosphere itself \citep{Pollack1996}, which might even delay runaway gas accretion \citep{Alibert2018, Venturini_2020_pla_chem, Guliera2020}, or when a large gaseous envelope has already formed \citep{Shibata2019, Shibata2020}. The accretion efficiency depends crucially on the size of the planetesimals \citep{Levison2010, Fortier2013, Johansen2019pla} and the migration speed of the planet \citep{TanakaIda1999}. Furthermore, small pebbles beyond the planetary orbit are subject to turbulent motions caused by the spiral arms of the giant planet; this can transport the pebbles to upper layers of the disk, where they can then subsequently be accreted by the giant planet \citep{Bi2021, Szulagyi2021} through a meridional flow around the planet \citep{Morbidelli2014}.

A detailed modeling of further solid accretion is beyond the scope of this work, and we only artificially added 30 Earth masses (Table~\ref{tab:add_solids}) of solids into the planetary atmospheres of our planets formed via pebble and gas accretion (Fig.~\ref{fig:growthtracks}). For simplicity, we used the solid composition at the final orbital position of the planet for the composition of the added material. While this approach is clearly simplified, it illustrates a very important effect: a change in the refractory content in the planetary atmosphere due to solid accretion. 

In Fig.~\ref{fig:owen_pla_eff} we show the atmospheric composition of the same planets as in Fig.~\ref{fig:growthtracks}, but with 30 Earth masses of solids added into the atmosphere. It is clear that the addition of solids into the atmosphere increases the refractory-to-volatile ratio in the atmosphere for all the different planets formed in our simulations\footnote{The planet starting at 3 AU in the disk with $\alpha=10^{-4}$ shows mostly an increase in oxygen, which is caused by the final position of the planet close to the water ice line, where the solid composition is dominated by water ice.}. Furthermore, the planetary C/O ratio decreases as well because of the oxygen bound in the refractory materials. 

We additionally note that the nitrogen abundance is not affected by the addition of further solids for the planets shown in Fig.~\ref{fig:owen_pla_eff}. This is related to the fact that no nitrogen is in solid form at the planetary position and can thus not be added by the accretion of further solids into the planetary atmosphere. This effect clearly illustrates the implications of our model: The volatile-to-refractory content in planetary atmospheres could give important constraints if solid accretion into atmospheres is efficient. 

\section{Jupiter and Saturn}
\label{sec:Jupgrowth}

To illustrate the effects of the accretion of vapor-rich material on the atmospheric composition of Jupiter and Saturn, we employed a simplified formation scenario in which the planets do not migrate from their current positions. We only modeled gas accretion onto the planetary cores, which we assumed to be fully formed at 1 Myr, in line with suggestions from cosmochemical studies \citep{Kruijer2017}.

\subsection{Simple growth model for Jupiter and Saturn}

\begin{figure}
    \centering
    \includegraphics[width=.45\textwidth]{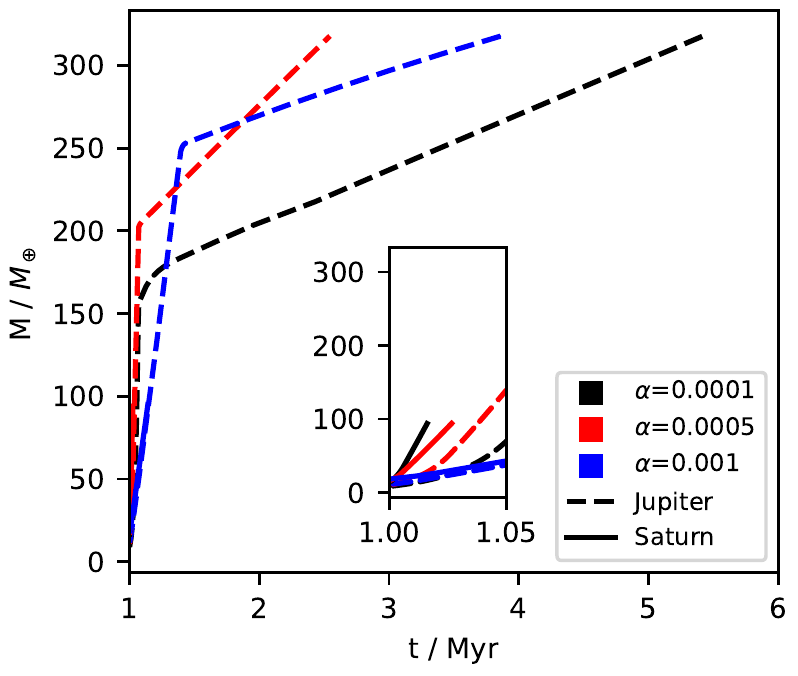}
    \caption{Mass growth of the gaseous components of Jupiter and Saturn as a function of time. We start our simulations with an already fully formed core at 1 Myr.}
    \label{fig:JupSatgrowth}
\end{figure}

We present here the ingredients for a simplified growth model of Jupiter and Saturn. We started under the assumption that the planetary core has already fully formed at 1 Myr, consistent with constraints from cosmochemistry \citep{Kruijer2017}. The planetary core mass corresponds to the pebble isolation mass at the current orbital positions of Jupiter and Saturn, and the planet is thus already in the gas accretion phase. We further took the very simplified approach that Jupiter and Saturn do not migrate during their evolution to illustrate the effects of vapor-enriched gas accretion on the atmospheric composition of these planets.

As the gas accretion rates in our nominal model are quite high (Fig.~\ref{fig:growthtracks}), we modeled the formation of Jupiter and Saturn in environments with a lower gas surface density because the gas surface density sets, alongside the viscosity, the gas accretion rate in the disk-limited regime \citep{Ndugu2021}. The gas disk masses are 0.0256, 0.0128, and 0.00256 M$_\odot$ for $\alpha=10^{-4}$, $5 \times 10^{-4}$, and $10^{-3}$, respectively, allowing the same gas delivery through the disk's accretion rate ($\dot{M}_{\rm gas} \propto \Sigma_{\rm g} \nu$) for all models. We stopped the integration once the masses of Jupiter and Saturn are reached. We show the growth of Jupiter and Saturn in Fig.~\ref{fig:JupSatgrowth}.

Initially, as the planets start to accrete gas, they can feed off the material inside the planet's horseshoe region \citep{CridaBitsch2017, Bergez2020}, allowing initially fast accretion. In fact, the amount of gas close to Saturn's position allows that Saturn accretes its gaseous envelope in less than 100 kyr. Jupiter, on the other hand, after it accretes all the material in its horseshoe region, feeds off the disk's gas supply. Due to the reduced gas surface density compared to our nominal model, Jupiter accretes gas for a few megayears after its core formation before it reaches its final mass. The time Jupiter needs to reach its final mass is nearly independent of the disk's viscosity, especially in the final stages of gas accretion, because the gas flow through the disk is the same in all three simulations. We note that the gas accretion time for Jupiter seems to be in line with the lifetime of the protosolar nebular \citep[e.g.,][]{Wang2017}, but obviously Saturn grows too quickly. We will address the combined growth (and migration) of Jupiter and Saturn in a future work.

\subsection{Implications for the formation of Jupiter and Saturn}

\begin{figure*}
    \centering
    \includegraphics[width=\textwidth]{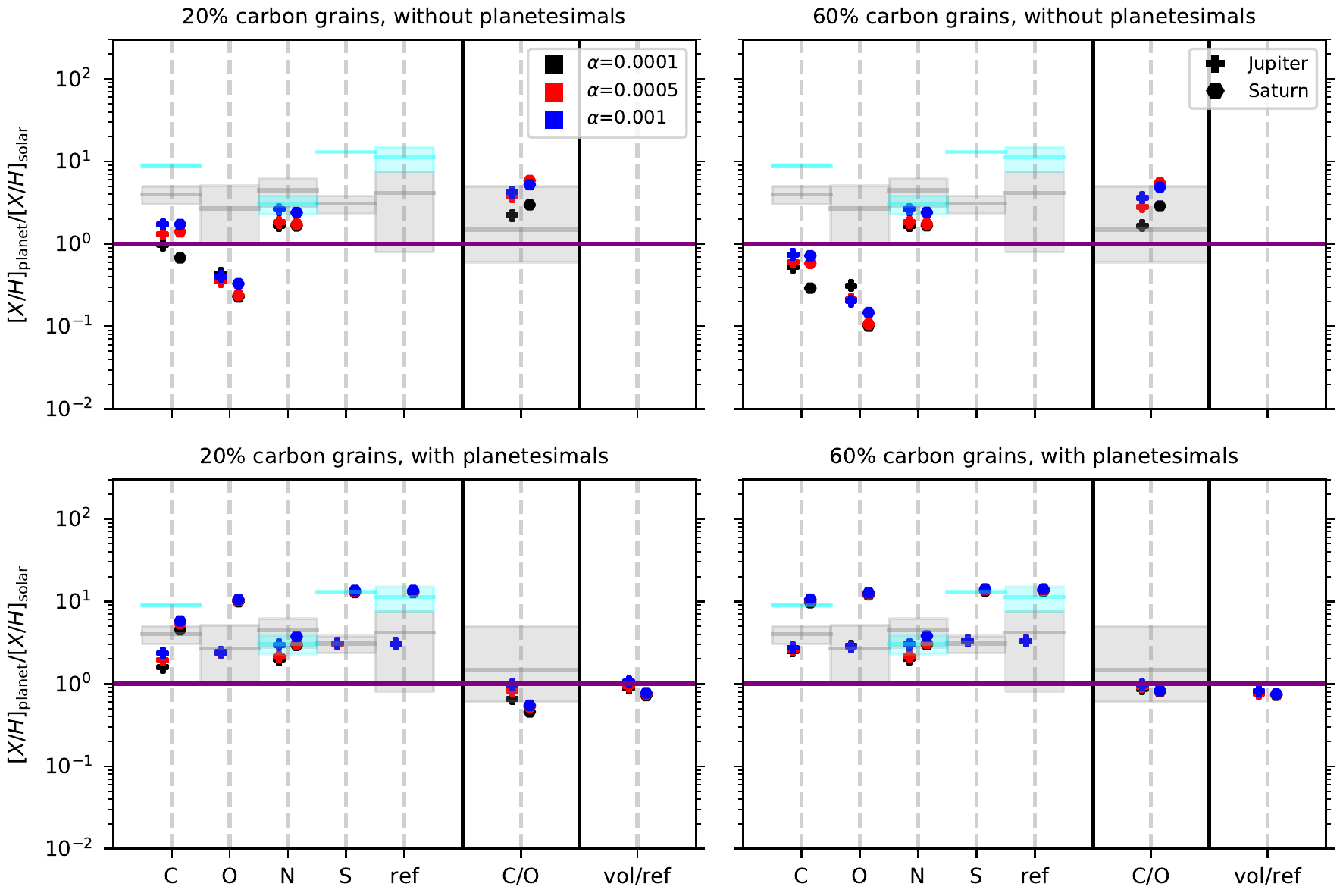}    
    \caption{Atmospheric compositions of Jupiter and Saturn from our model and the corresponding observations \citep{Atreya2016, Li2020Jup}. The left panels display models with 20\% of the carbon abundance locked in refractory carbon grains, whereas the right panels show models with 60\% locked in carbon grains. The bottom panels display the effect of additional solids added to the atmosphere (see Table \ref{tab:add_solids}). The gray bands correspond to measurements in Jupiter's atmosphere, while the horizontal light blue line marks the measurements for Saturn. We show here only the atmospheric abundances without mixing of the planetary core into the atmosphere because Jupiter has a core \citep{Wahl2017}. The refractory content is zero if we do not add any additional solids because our model does not allow the accretion of solids during the gas accretion phase.}
    \label{fig:JupSat}
\end{figure*}

Some elemental abundances have been constrained from previous missions to Jupiter  (for a recent review, see \citealt{Atreya2016}), and the recent Juno mission revealed the oxygen abundance around Jupiter's equator \citep{Li2020Jup}. Under the assumption that the oxygen abundance around Jupiter's equator corresponds to the bulk composition, we determined Jupiter's C/O, as marked in Fig.~\ref{fig:JupSat}, which shows a super-solar C/O. We discuss now the implications of our model on the C, O, N, and refractory abundances of Jupiter.

In order to achieve a super-solar C/O, the planet needs to accrete gas enriched with carbon, which is easiest to achieve beyond the water ice line \citep{Oeberg2011, Schneider2021I}. While the C/O ratio is matched quite well, the actual O/H values are too low compared to that of Jupiter (Jupiter's O/H is $2.7^{+2.4}_{-1.7}$ solar; \citealt{Li2020Jup}) because Jupiter stays beyond the water ice line at all times in our model. The addition of oxygen-rich solids (e.g., planetesimals, comets) could provide more oxygen to Jupiter, or an inward migration followed by an outward migration, as in the grand tack scenario \citep{Walsh2011}, could allow the accretion of more oxygen from the gas phase of the protoplanetary disk (see also Fig.~\ref{fig:owen}). The measured C/H of Jupiter is larger than in our simulations, which could also be increased by the accretion of further carbon-containing solids or by gas accretion close to evaporation fronts of carbon species.

Recent studies \citep{Ali-Dib2017, Oeberg2019, Bosman2019} put forward the idea that Jupiter formed in the outer regions of the disk and then migrated inward \citep{Bitsch2015} due to the nitrogen and noble gas abundances in Jupiter's atmosphere, which are enriched by a factor of of two to four compared to the Sun \citep{Owen1999}. The underlying idea of these works is that nitrogen and the noble gases were frozen out and, as such, could be accreted in solid form. This idea is also based on the large abundance of N$_2$ compared to NH$_3$ \citep{Boogert2015, Cleeves2018, Pontoppida2019}. However, NH$_3$ ices could be locked in salts \citep{Altwegg2020Comet}, allowing an accretion of nitrogen via solids even closer to the host star.

Within our model, we observe super-solar nitrogen abundances for Jupiter and Saturn (Fig.~\ref{fig:JupSat}). This is caused by the evaporation of inward drifting nitrogen-rich pebbles, which enriches the gas phase and consequently the planetary atmosphere. The same should apply for noble gases with very low evaporation temperatures. Our model thus implies that Jupiter did not necessarily need to form close to the N$_2$ evaporation front.

For Saturn, only constraints for the C/H ratio are available \citep{Atreya2016}. Our solid-free model clearly underproduces the C/H ratio of Saturn, implying that pebble evaporation alone can probably not explain Saturn's composition. On the other hand, our simulations are very simplistic regarding the interior distribution of material. Studies of Jupiter have clearly revealed a diffused core and metallicity gradient inside the planet \citep{Wahl2017, Vazan2018, Debras2019}, which could enhance the elemental abundances compared to our simple model (see the mixing of the core in Fig.~\ref{fig:owen}).

In Fig.~\ref{fig:JupSat} we also show the refractory content in the planetary atmospheres, which is  by construction zero for the models without additional solids because our model does not allow the accretion of solids during the gas accretion phase. The refractory content of Jupiter is still unknown for most elements; only the sulfur and phosphor abundances are measured ($\approx$ 4 $\times$ solar; \citealt{Atreya2016}). However, a small sulfur fraction is in volatile form as H$_2$S, which could allow sulfur accretion via the gas phase if the planet migrates into the inner disk (Fig.~\ref{fig:owen}), as proposed in the grand tack scenario \citep{Walsh2011}. On the other hand, most of the phosphor is locked in refractories \citep[e.g.,][]{Lodders2003}, making this scenario very unlikely.

We also demonstrate how the accretion of additional solids (see Table \ref{tab:add_solids}) would enrich the planetary atmosphere with refractories (see the bottom panels of Fig.~\ref{fig:JupSat}), as discussed in Sect.~\ref{sec:solids}. We calibrated the amount of additionally accreted solids to match the measure sulfur abundance in the atmospheres of Jupiter and Saturn. This corresponds to an additional amount of 10-15 Earth masses of solids that needs to be accreted. However, it is clear from our simplified model that pebble drift and evaporation is an important ingredient for the volatile content of the atmospheres of Jupiter and Saturn, especially for the nitrogen (and noble gas) content.

\section{Conclusion}\label{sec:conclusion}

We have studied the influence of pebble evaporation on the atmospheric composition of giant planets. In particular, our simulations show that pebble evaporation results in a significant enhancement of the volatile content in the planetary atmospheres compared to the refractory contents. This is caused by accretion of gas enriched in volatiles due to pebble evaporation, which could also explain the C/H and C/O ratios of $\tau$ Boo b \citep{Pelletier2021}. Our exact results depend crucially on the underlying chemical model, which determines how much of a given element is in volatile or refractory form (see Sect.~\ref{sec:methods}); however, independently of the exact chemical model, pebble evaporation plays a crucial role in determining the volatile content in a giant planet's atmosphere.

Our simulations show that the C/O ratio alone might not be enough to constrain the formation history of giant planets, and additional constraints, either from other elements \citep[see also][]{Turrini2021} or from direct abundances \citep[see also][]{Notsu2020}, are needed. Furthermore, our simulations show that Jupiter's nitrogen enrichment could be caused by the accretion of nitrogen-rich gas, implying that Jupiter did not need to form in the very outer regions of the Solar System as proposed \citep{Oeberg2019, Bosman2019}.

If additional solids are added to the planetary atmosphere, the refractory content in the planetary atmosphere increases, reducing the volatile-to-refractory fraction. Our simple model also indicates that the addition of refractory material into Jupiter's atmosphere is needed to match the observational constraints. We note, however, that this is also influenced by our model assumption that does not allow the accretion of refractories during the gas accretion phase. Therefore, our study suggests that a large refractory content in planetary atmospheres might be a sign of additional solid pollution (via planetesimals, giant impacts, or dust transported through meridional flows during the gas-disk phase), as also speculated in \citep{Lothringer2021}. Future observations of giant planet atmospheres could thus help to constrain the efficiency of solid accretion into atmospheres.

\begin{acknowledgements}
A.D.S and B.B. thank the European Research Council (ERC Starting Grant 757448-PAMDORA) for their financial support. A.D.S. acknowledges funding from the European Union H2020-MSCA-ITN-2019 under Grant no. 860470(CHAMELEON) and from the Novo Nordisk Foundation Interdisciplinary Synergy Program grant no. NNF19OC0057374. We thank the anonymous referee for the useful remarks that helped improve the manuscript.
\end{acknowledgements}

\begingroup
\bibliographystyle{aa}
\bibliography{ms}
\endgroup

\newpage
\begin{appendix}

\section{Disk}\label{sec:disk}
\begin{figure*}
    \centering
    \includegraphics[width=\textwidth]{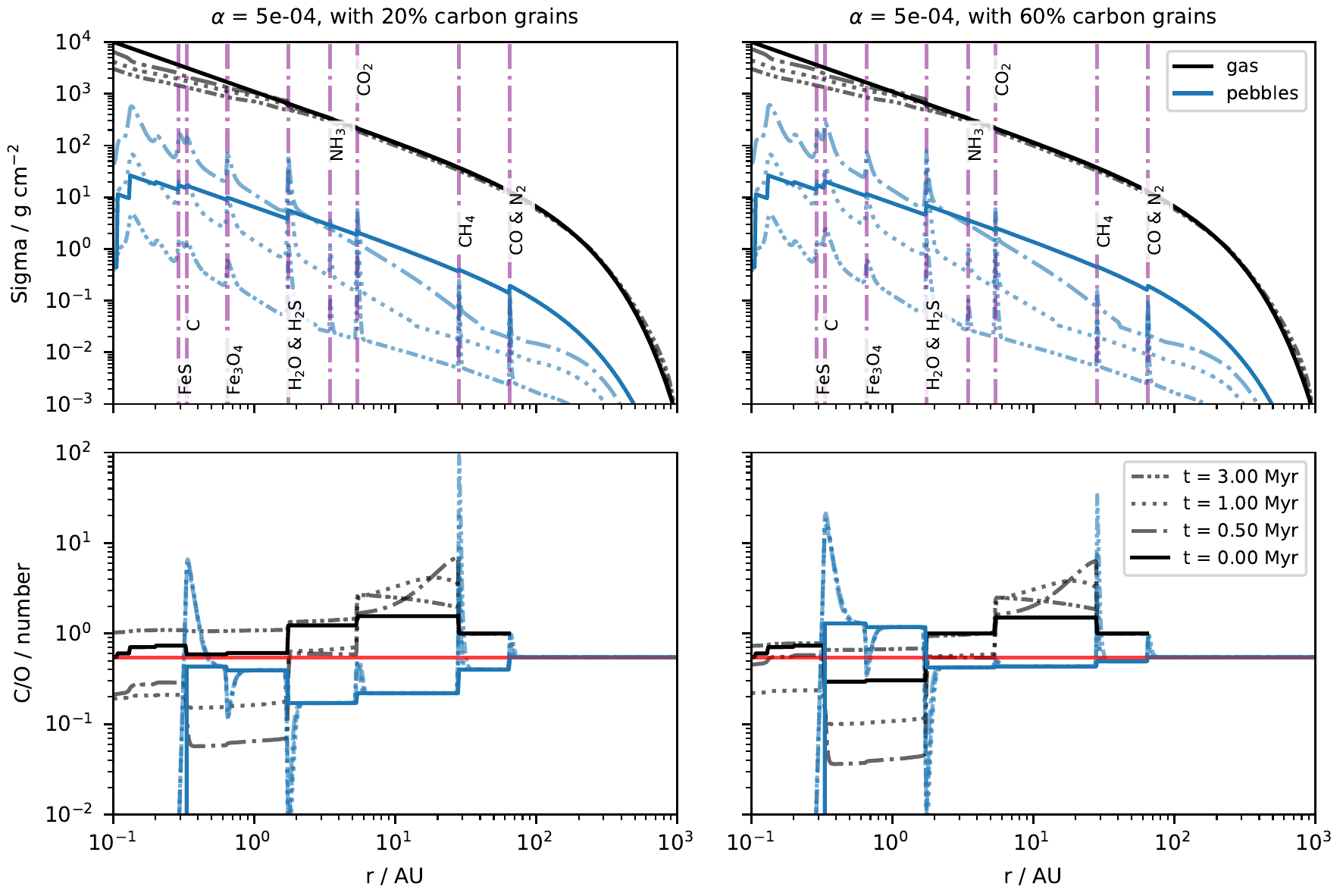}
    \caption{Disk evolution (top) for $\alpha=\num{5e-4}$ and the corresponding C/O ratio in gas and pebbles (bottom). The left panels display models with 20\% of the carbon abundance locked in refractory carbon grains, whereas the right panels show models with 60\% locked in carbon grains. The red line in the bottom panel marks the solar C/O.}
    \label{fig:sigma}
\end{figure*}
In Fig.~\ref{fig:sigma} we show the evolution of a disk with $\alpha=\num{5e-4}$ and the corresponding C/O ratios for the two different refractory carbon contents used in our model. The overall evolution of the gas and pebble surface densities is very similar compared to our nominal model \citep{Schneider2021I}, the main difference being the reduction in the pebble pileups at the CO, CH$_4$, and CO$_2$ evaporation fronts due to less available CO, CH$_4$, and CO$_2$ (Table~\ref{tab:comp}). On the other hand, we now observe a larger pebble pileup around the carbon grain evaporation front in the inner disk regions as well as an increased pileup around the water ice line because less oxygen is bound in CO and CO$_2$, increasing the water abundance.

Comparing the two models, the corresponding C/O ratios in the pebbles and gas phase differ mostly in the inner disk regions. In fact, the C/O of the solids and the gas interior to the water ice line (up until the carbon grains evaporate) is now super-solar and subsolar, respectively, opposite to the nominal model \citep{Schneider2021I}. This is caused by the fact that now large amounts of carbon grains can exist in the inner disk regions. Furthermore, the large spike in the solid C/O ratio at the methane ice line is significantly reduced due to the lower methane fraction. The evolution of the C/O ratio in time shows the effects described in \citet{Schneider2021I}, namely the competition between fast inward drifting pebbles and slow vapor diffusion, resulting in a change in the C/O ratio of solids and gas in time.

\end{appendix}

\end{document}